\documentclass[prl,twocolumn,amsmath,amssymb]{revtex4}

\usepackage{graphicx}
\usepackage{upgreek}

\begin{document}

\title{Extremely confined gap surface plasmon modes excited by electrons}

\author{S{\o}ren~Raza,$^{1,2,}$\footnote{Authors contributed equally} Nicolas~Stenger,$^{1,3,*}$ Anders~Pors,$^{4,*}$ Tobias~Holmgaard,$^5$ Shima~Kadkhodazadeh,$^2$ Jakob~B.~Wagner,$^2$ Kjeld~Pedersen,$^5$ Martijn~Wubs,$^{1}$ Sergey~I.~Bozhevolnyi,$^{4}$ \& N.~Asger~Mortensen$^{1,3}$}

\affiliation{$^1$Department of Photonics Engineering, Technical University of Denmark, DK-2800 Kgs. Lyngby, Denmark\\
$^2$Center for Electron Nanoscopy, Technical University of Denmark, DK-2800 Kgs. Lyngby, Denmark\\
$^3$Center for Nanostructured Graphene (CNG), Technical University of Denmark, DK-2800 Kgs. Lyngby, Denmark\\
$^4$Department of Technology and Innovation, University of Southern Denmark, DK-5230 Odense M, Denmark\\
$^5$Department of Physics and Nanotechnology, Aalborg University, DK-9220 Aalborg {\O}st, Denmark}

\begin{abstract}
High spatial and energy resolution EELS can be employed for detailed characterization of both localized and propagating surface plasmon excitations supported by metal nanostructures, giving insight into fundamental physical phenomena involved in various plasmonic effects. Here, applying EELS to ultra-sharp convex grooves in gold, we directly probe extremely confined gap surface plasmon (GSP) modes excited by swift electrons in nanometer-wide gaps. Both experimental and theoretical EELS data reveal the resonance behavior associated with the excitation of the antisymmetric (with respect to the transverse electric-field component) GSP mode for extremely small gap widths, down to $\sim 5$~nm. It is argued that the excitation of this mode, featuring very strong absorption, plays a crucial role in the experimental realization of non-resonant light absorption by ultra-sharp convex grooves with fabrication-induced asymmetry. Occurrence of the antisymmetric GSP mode along with the fundamental GSP mode exploited in plasmonic waveguides with extreme light confinement is a very important factor that should be taken into account in the design of plasmonic nanophotonic circuits and devices.
\end{abstract}

\maketitle


Surface plasmons (SPs), i.e., the collective excitation of the conduction electrons localized at the metal surface, allow for subwavelength localization and guiding of light~\cite{Gramotnev:2010}, and electric field enhancements on the nanoscale~\cite{Hao:2004}. These attractive properties of the SPs have found application in a wide variety of fields; from medical applications, such as bio-sensing~\cite{VazquezMena:2011} and cancer therapy~\cite{Lal:2008}, to plasmonic waveguiding~\cite{Bozhevolnyi:2006b}. However, the experimental mapping of SPs on the nanoscale using light still remains an inherently difficult task due to the diffraction limit. In contrast, the use of swift electrons to excite SPs offers a powerful spectroscopic technique known as electron energy-loss spectroscopy (EELS)~\cite{GarciadeAbajo:2010}.

\begin{figure}[b!]
\centering
\includegraphics[scale=0.95]{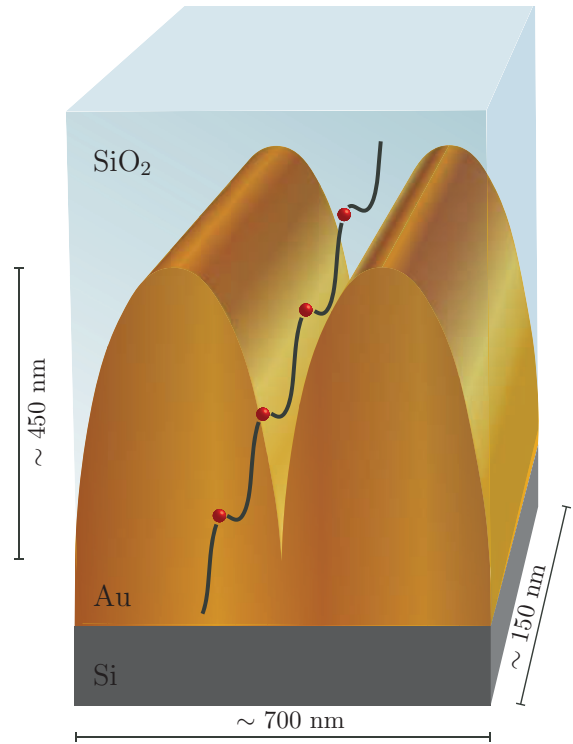}
\caption{Artistic impression of a single gold nanogroove with the swift electron beam moving parallel to the groove axis. The groove is filled with silicon dioxide and the substrate is silicon. The period and height of the grooves are determined from the STEM images, while the thickness of the sample has been inspected in a SEM and also estimated from EELS data using both the log-ratio method and Kramers--Kronig analysis~\cite{EgertonBook:2011}.\label{fig:fig0}}
\end{figure}

\begin{figure*}[t!]
\centering
\includegraphics[scale=1.5]{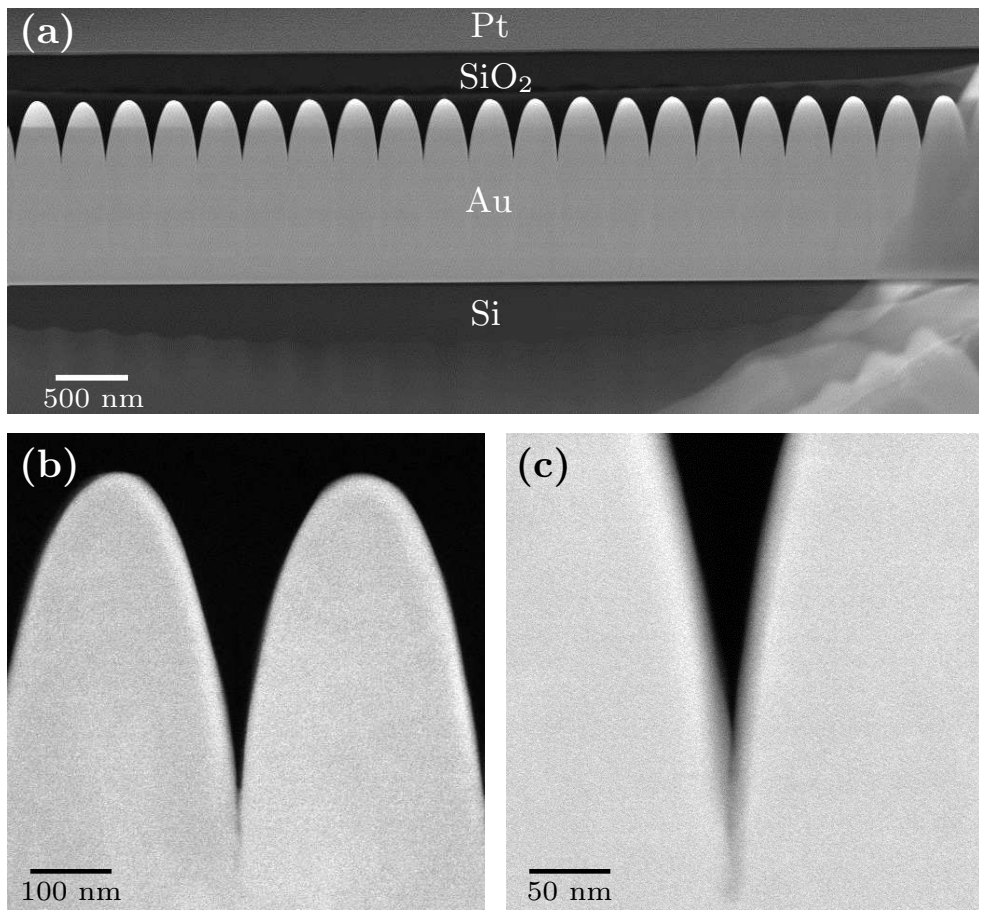}
\caption{STEM micrographs of a 1D gold nanogroove. STEM images of \textbf{(a)} sample overview with material labels, \textbf{(b)} single groove zoom-in and \textbf{(c)} ultra-sharp groove crevice. The STEM images display the periodicity of the structure and the similarity in shape of each groove. Furthermore, the grooves are quite symmetric along the center line [dashed line in \textbf{(b)}] and extremely sharp, with around 5~nm gap sizes in the crevice.\label{fig:fig1}}
\end{figure*}

Performing EELS in state-of-the-art transmission electron microscopes (TEMs), equipped with a monochromator and aberration corrector, offers unmatched simultaneous spatial and spectral resolution~\cite{GarciadeAbajo:2010,Nicoletti:2013}. Since Ritchie~\cite{Ritchie:1957} predicted the excitation of SPs using electrons, low-loss EELS has become an increasingly popular technique in plasmonics, especially to study and map localized SP resonances~\cite{Ouyang:1992,Nelayah:2007,Bosman:2007,Koh:2009,Scholl:2012,Raza:2013,Nicoletti:2013}. While the major part of plasmonic EELS studies have been focused on localized SP resonances, EELS has also been used to study propagating SPs (i.e., waveguide modes) in e.g. metal thin films~\cite{Pettit:1975} and nanowires~\cite{Nicoletti:2011,Rossouw:2013}. However, gap SP (GSP) modes, i.e., the propagating SP modes in a dielectric gap between two metals~\cite{Prade:1991}, has to our knowledge never been experimentally examined with EELS. The GSP modes occur in a variety of geometries, from the simplest 1D metal-insulator-metal (MIM) waveguide~\cite{Economou:1969a} to more advanced structures such as the convex groove, V-groove, trench and stripe waveguides~\cite{Sondergaard:2012,Bozhevolnyi:2008a}. Furthermore, the GSP modes offer enhanced properties compared to the usual propagating SP mode, such as extreme light confinement with improved propagation distances~\cite{Miyazaki:2006,Dionne:2006a}, negative refraction~\cite{Shin:2006,Lezec:2007} and highly efficient light absorption~\cite{Sondergaard:2012}, to name a few important and distinct features of the GSP modes in plasmonics.

In this work we report on the first experimental study of GSPs in ultra-sharp gold convex nanogrooves using EELS. The convex nanogroove is ideal for studying the GSP mode, since the width of the insulating layer (gap size) decreases as the position of the electron probe is moved down the nanogroove (see Fig.~\ref{fig:fig0}), allowing us to map the evolution of the GSP mode for varying gap size in a single groove. We find that the antisymmetric GSP mode, which is excited by the electrons, increases in energy as the gap size decreases, and additionally we observe that the antisymmetric GSP is still present at an extremely narrow gap of only 5~nm. Furthermore, we argue why the excitation of this mode, featuring very strong absorption, plays a crucial role in the experimental realization of non-resonant light absorption by ultra-sharp convex grooves with fabrication-induced asymmetry~\cite{Sondergaard:2012}.

The fabrication of the gold nanogrooves is done using a focused ion beam (FIB) setup. A 1.8-$\upmu$m-thick gold film is deposited on a flat silicon substrate, whereafter areas of the gold are milled by the FIB to create the groove structure. With this technique ultra-sharp grooves with a 1D period of approximately 350~nm are fabricated.  In the FIB, a layer of silicon dioxide is then deposited to separate the grooves from the top platinum layer used for protecting the sample during the preparation of the TEM lamella. The thickness of the silicon dioxide layer (approximately 500~nm) is enough to avoid the influence of the platinum layer, when performing EELS measurements inside the groove. With a micromanipulator the nanogroove sample is positioned on the TEM lift-out grid such that the electron beam passes perpendicularly to the section of the sample and parallel to the axis of the groove, as illustrated in Fig.~\ref{fig:fig0}. In order to characterize the grooves in the TEM with EELS, we use the FIB to thin the nanogrooves to a thickness of approximately 150~nm (in the direction parallel to the axis of the groove). This ensures that the sample is sufficiently transparent for the electron beam, while at the same time mechanically stable. Note that as a consequence of the thinned sample, the structure can only be considered as a waveguide, facilitating propagating modes, when the electron beam probes (local) GSP modes in the depth of the grooves, i.e., when the groove width is considerably smaller than the sample thickness of $\sim150$~nm.

Figure~\ref{fig:fig1} shows examples of scanning TEM (STEM) images of the gold nanogroove sample. In Fig.~\ref{fig:fig1}(a), an overview image of the sample is displayed with the gold nanogroove on top of a flat silicon substrate. The grooves are filled with silicon dioxide and the top platinum layer can also be seen. Figure~\ref{fig:fig1}(b) shows a zoom-in of a single groove, displaying the almost perfect symmetry with respect to the middle of the groove [white line in Fig.~\ref{fig:fig1}(b)]. However, as we will discuss later, the slight geometric asymmetry of the groove will be crucial in understanding the plasmonic black gold effect studied in Ref.~\onlinecite{Sondergaard:2012}. Finally, Fig.~\ref{fig:fig1}(c) is a close-up of a groove crevice, showing the extreme sharpness of the groove.  The side-to-side width of the groove from the top to the bottom is calculated with an in-house written image analysis code and ranges from 320~nm down to sizes smaller than 5~nm, thus confirming the ultra-sharp shape of the grooves. While slight fluctuations in shape and groove depths are seen, the grooves are overall impressively similar (which is also reflected in the subsequent EELS measurements).

To characterize the grooves we have used an aberration-corrected FEI Titan STEM operated at an acceleration voltage of 300~kV (corresponding to an electron velocity of $v=0.776c$) and with an electron probe size (i.e., FWHM probe profile) of approximately 5~{\AA}. The system is equipped with a monochromator thus allowing us to perform EELS measurements with an energy resolution of $0.15\pm0.05$~eV. We performed a detailed analysis of six grooves on the same sample by systematically collecting EELS data from the top to the bottom of the groove (Supplemental Material). Since the results obtained for these six grooves are very similar (see Supplementary Figure~1), we only present results for two of the grooves.

\begin{figure}
\centering
\includegraphics[scale=1]{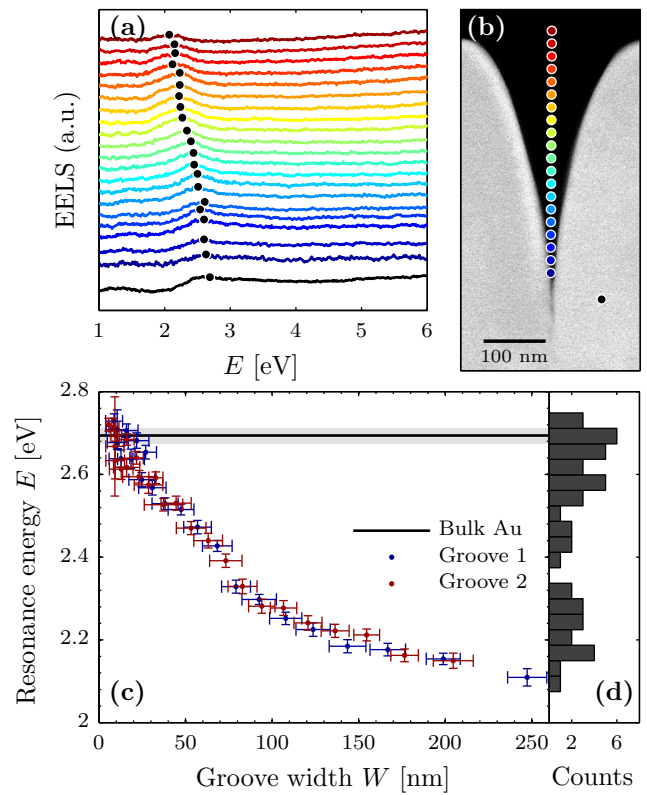}
\caption{Experimental EELS data. \textbf{(a)} Waterfall plot of experimental EELS measurements at the corresponding positions indicated on the groove image in \textbf{(b)}. The bottom black line displays the experimental EELS spectrum of bulk Au. \textbf{(c)} Peak resonance energy as a function of groove width for two different grooves, along with the measured bulk resonance energy of Au (solid black line with grey area displaying the error bar). \textbf{(d)} Histogram displaying the number of resonance energies within bin intervals of $0.04$~eV [projection of data in \textbf{(c)} onto the $E$-axis]. The bin interval is chosen as the average energy error bar size.\label{fig:fig2}}
\end{figure}

The EELS data along with their corresponding electron probe positions in the groove are displayed in Figs.~\ref{fig:fig2}(a-b). The only data treatment of the EELS data has been to subtract the zero-loss peak from the raw data using the reflected-tail method. The EELS data are relatively featureless for a broad range of energies, but do show clear resonance peaks due to the excitation of SP mode(s). As the most prominent feature, we observe that the resonance peak blueshifts from 2.1~eV to approximately 2.6~eV when the position of the electron probe is moved from the top towards the bottom of the groove. This peak position sensitivity to the groove width, especially apparent for small groove widths, is a clear indication that\textit{ local} modes rather than (global) groove modes are probed. Moreover, the existence of local modes in the crevice of the groove suggests the excitation of GSP modes related to metal-insulator-metal (MIM) configurations, as we argue in detail below.

The MIM waveguide supports two fundamental GSP modes, a symmetric GSP (sGSP) mode and an antisymmetric GSP (aGSP) mode~\cite{Economou:1969a,Raza:2013b}, where the symmetry nomenclature of the modes refer to the symmetry of the transversal component of the electric field. For realistic MIM configurations with subwavelength dielectric gaps, the aGSP mode is considerably more lossy than the sGSP mode~\cite{Dovoyan:2010}. When the electron probe is positioned in the center line of the groove, the negatively-charged electrons induce positive charges on each metal surface of the groove, thereby exciting GSP modes that sustain a symmetrical charge distribution. For convenience, we now define the $z$-axis to be along the electron path, while the $xy$-plane is orthogonal to the electron path, see the inset of Fig.~\ref{fig:fig1}(c). The GSP modes excited in EELS are therefore symmetric with respect to the $z$-component (longitudinal component) of their electric field, and antisymmetric with respect to the $x$- and $y$-components (transversal components). Accordingly, the local GSP mode excited by the electron probe is labeled an aGSP mode. Note that the analogy between local aGSP modes in the groove and in MIM waveguides is supported by our simulations (below) showing the antisymmetric MIM GSP mode blueshifting its resonance with decreasing gap width.

In Fig.~\ref{fig:fig2}(c) we plot the energy of the resonance peak $E$ as a function of the width $W$ of the groove for two different grooves. The plot shows first a slow increase of the resonance energy from 2.1~eV to 2.3~eV when the groove width decreases from 250~nm down to 100~nm. This behavior is then followed by a stronger blueshift from 2.3~eV to 2.6~eV for widths decreasing from 100~nm down to 5~nm. As numerically calculated EELS data of groove waveguides (to be discussed below) display the same trend, we interpret the dependence $E(W)$ as a result of two (spectrally-close) modes excited simultaneously but with different strengths depending on the position of the electron probe. For widths $W\gtrsim100$~nm, the local aGSP mode is weakly excited due to the increased distance between the electron and the metal-insulator interfaces, which therefore suggests the excitation of localized SPs supported by the top of the grooves. Accordingly, the slow increase in resonance energy (when the groove width decreases from 250~nm down to 100~nm) represents the transition from localized SP excitations to a MIM description with propagating local aGSP modes. In the case of groove widths $W\lesssim100$~nm, on the other hand, the local aGSP mode dominates the EELS data, signified by the strong dependence of the resonance peak on the groove width.
Notably, the aGSP resonance energy in the crevice is very close to the measured bulk resonance energy of gold of 2.7~eV, also shown in Fig.~\ref{fig:fig2}(c) as a black solid line. For extremely narrow groove widths ($<10$~nm) the electron will also partly penetrate the gold groove walls. This is in part due to the surface roughness of the walls, but also due to the convergence angle of the focused electron probe. The convergence semi-angle is approximately 16~mrad, which means that the electron trajectory is displaced from the straight-line path by around 2.5~nm at the exit of the groove (under the assumption that the focus point of the beam is at the front plane of the groove). Thus, at very narrow widths there is a possibility of both GSPs and bulk plasmons being excited. Due to the energy resolution of the microscope ($0.15\pm0.05$~eV), it therefore becomes increasingly difficult to distinguish between the aGSP resonance energy (at 2.6~eV) and the resonance energy of bulk gold (at 2.7~eV) in the EELS data. We point out that we can not distinguish the two resonance peaks from a single spectrum, since their difference in resonance energy is below the resolution of the microscope. However, depending on the exact position of the electron probe, we excite one resonance more efficiently than the other, allowing us to determine the energy of that particular resonance. This effect is visible in the spread of the resonance energies for narrow widths in Fig.~\ref{fig:fig2}(c) and is also confirmed in the histogram in Fig.~\ref{fig:fig2}(d). The histogram represents the measured resonance energies projected along the energy axis and binned into energy intervals of 0.04~eV. The histogram shows a larger number of counts both at the resonance energy of the aGSP mode (at 2.6~eV) and at the bulk resonance energy of gold (at 2.7~eV), with a dip in between these energies, thus supporting the interpretation that two different resonances close in energy are present.

The two different grooves in Fig. \ref{fig:fig2}(c) show almost identical trends for the SP resonances, indicating that the shape variation from groove to groove is small. More astonishing is that both grooves support local aGSPs in extremely narrow gaps of only 5~nm. The two energy-shift regions, i.e., the slow increase for $W\gtrsim100$~nm and the faster increase for $W\lesssim100$~nm, and the presence of the aGSP close to the bulk resonance energy in ultra-narrow gaps are also observed for all other grooves studied during this work (see Supplementary Figure 1).

\begin{figure}
\centering
\includegraphics[scale=1]{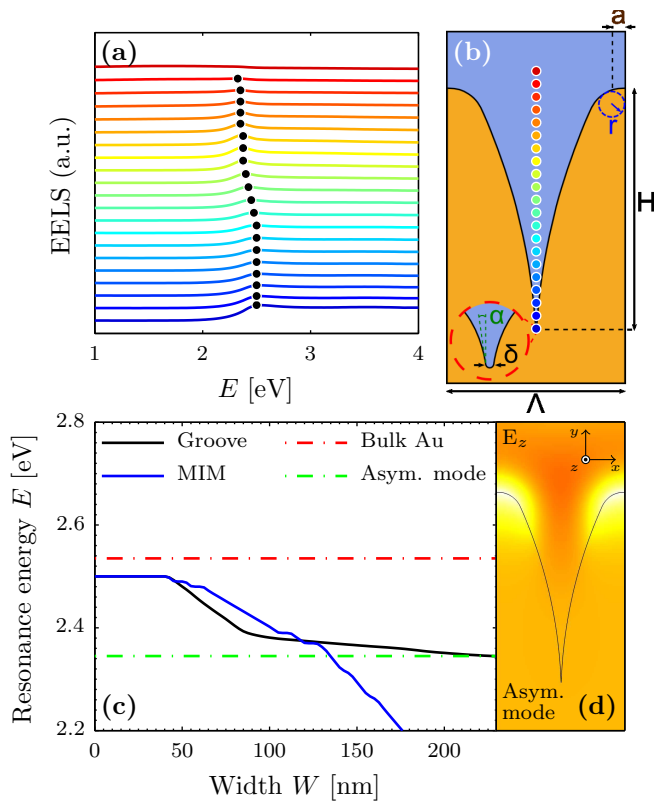}
\caption{Theoretical EELS data. \textbf{(a)} Waterfall plot of theoretical EELS calculations at the corresponding position indicated on the groove model in \textbf{(b)}. \textbf{(c)} Longitudinal component of the scattered electric field (total E-field subtracted the E-field of the electron) for an electron beam probing at the resonance energy of four different groove widths $W$. \textbf{(d)} Peak resonance energy determined from EELS calculations as a function of width for the groove model in \textbf{(b)} (black line), MIM (Au-SiO$_2$-Au) waveguide (blue line), and bulk gold (red dash-dotted line). The green dash-dotted line represents the resonance energy of the antisymmetric mode supported by the groove. \textbf{(e)} Longitudinal component of the electric field of the antisymmetric groove mode. Permittivity for gold is taken from Ref.~\onlinecite{Johnson:1972}. We use $\varepsilon_{\text{SiO}_2}=2.1$.\label{fig:fig3}}
\end{figure}

To support the interpretation of the experimental EELS data, we have performed fully retarded EELS calculations using the commercial software COMSOL Multiphysics. For the numerical analysis we consider the nanogroove structure to be infinite in the direction of the electron beam i.e., $z$-direction, which allows us to simplify the problem to 2D.  Although the finite thickness of the sample is thus neglected, we can still expect that the 2D approximation is accurate for narrow groove widths (which is the region of interest) where the width-to-sample thickness ratio is low. Furthermore, we assume that the groove has perfect mirror symmetry, thereby neglecting any slight geometric asymmetries present in the sample.

The results of the theoretical analysis are summarized in Fig.~\ref{fig:fig3}. Figure~\ref{fig:fig3}(a) shows calculated EELS data at the electron beam positions indicated in Fig.~\ref{fig:fig3}(b), displaying flat spectra with distinct resonances visible, in accordance with our experimental measurements. Furthermore, the same blueshift trend of the  resonance peak shows up in our simulations when the position of the electron probe is moved towards the bottom of the groove. The SP resonance shifts from approximately 2.3~eV to 2.5~eV. In Fig.~\ref{fig:fig3}(d) the black line displays the calculated resonance energy as a function of the groove width. We see that the initial slow blueshift for $W \gtrsim 100$~nm followed by a steep blueshift for $W \lesssim 100$~nm is accurately captured in our theoretical model. Furthermore, for $W\lesssim 50$~nm the calculations show a plateau for the resonance energy at $2.5$~eV, which is close to the bulk resonance energy of gold (red dash-dotted line). This is again in excellent agreement with our experimental observations, although the bulk resonance energy of gold from the data in Ref.~\onlinecite{Johnson:1972} is slightly lower than the bulk resonance energy of our sample [at 2.7~eV, see Fig.~\ref{fig:fig2}(c)]. Figure~\ref{fig:fig3}(c) displays the longitudinal component of the scattered E-field for an electron beam probing at the resonance energy of four different groove widths, demonstrating that a local aGSP mode (with a local field distribution) is indeed excited for widths $W\lesssim100$~nm. In contrast, we see a noticeable field distribution at the top of the groove for $W\gtrsim 100$~nm, indicating the excitation of the groove-supported (global) antisymmetric plasmonic waveguide mode whose mode profile is shown in Fig.~\ref{fig:fig3}(e). This interpretation is further substantiated by the fact that the energy of the EELS resonance peak in Fig.~\ref{fig:fig3}(d) approaches the energy of the groove-supported antisymmetric mode (green dash-dotted line) for large groove widths. Summarizing, the peculiar dispersion of the single peak observed in the theoretical EELS data originates from the strong excitation of local aGSP modes in the bottom of the groove, with decreasing strength as the electron probe moves up the groove, while at the same time the excitation efficiency of the groove-supported antisymmetric mode increases. In the experimental setup with a sample thickness of $\sim 150$~nm, the groove-supported antisymmetric mode does not exist, but instead localized SP modes at the top of the groove are present in the EELS data.

To test the analogy between the local aGSP mode in our grooves and the corresponding mode in MIM waveguides, we have also performed EELS calculations on the MIM (Au-SiO$_2$-Au) waveguide. As in the simulations of the groove, we position the electron probe in the center of the MIM waveguide and calculate the resonance energy of the aGSP mode as a function of the width of the MIM waveguide (i.e. width of the insulating layer). The blue line in Fig.~\ref{fig:fig3}(d) displays the result. We see that the local aGSP mode in the groove is well-described for $W\lesssim125$~nm by the MIM aGSP mode, and is identical for $W\lesssim50$~nm, confirming that the two aGSP modes are indeed very similar. This is quite surprising as it seems that the ultra-sharp crevice of the groove does not alter the properties of the MIM aGSP mode significantly. From dispersion relation calculations of the MIM waveguide (not shown), we also find that the propagation length of the aGSP mode at the resonance energy (2.5~eV) is below 10~nm for a MIM waveguide with a width of 50~nm. Furthermore, the propagation length decreases with decreasing width, thus supporting the validity of the 2D approximation in our calculations when in the bottom of the groove.

When comparing the experimental EELS measurements with the theoretical groove simulations, we find that the experimentally measured aGSP resonance energy spans a broader range (2.1-2.6~eV) compared to the simulations (2.3-2.5~eV). As argued earlier, we attribute the discrepancy at narrow widths ($W\lesssim50$~nm) to the difference in the permittivity of gold of Ref.~\onlinecite{Johnson:1972} used in the theoretical calculations and the permittivity of the gold in our sample. To verify this statement, we have, from our EELS data, determined the permittivity of the gold in our sample using Kramers--Kronig analysis~\cite{EgertonBook:2011} (see Supplementary Figure~2). We find that the gold in our sample is less metallic [less-negative Re($\varepsilon$)] and more lossy [larger Im($\varepsilon$)] in the energy range of interest (2-2.7~eV) than the results by Ref.~\onlinecite{Johnson:1972}. While the exact mechanism for this alteration of the gold in our sample is not completely clear, we believe this may be related to gallium contamination and surface amorphization of gold from the procedure of thinning the sample using the FIB. Such FIB-related damages can alter up to few tens of nanometers of each of the groove surfaces, depending on the FIB conditions used~\cite{McCaffrey:2001,Kato:2004}. Note that despite the fact that the (probable) FIB-related damages affect the gold permittivity in the whole energy range considered (Supplementary Figure~2), the discrepancy between measurements and simulations at large groove widths ($W\gtrsim100$~nm) is in any case expected as the EELS peaks are related to physically different phenomena (localized SP and antisymmetric waveguide mode, respectively).

As an consequence of our EELS study, we can now add a crucial point to the discussion why fabricated ultra-sharp convex grooves in gold absorb light more efficiently than theoretically predicted~\cite{Sondergaard:2012}. The nanogroove structure has recently been shown to result in very efficient absorption of light. The incident light, which propagates downwards (i.e., in the $xy$-plane) for this purpose, couples through scattering off the groove wedges to the symmetric MIM GSP mode (not probed by our EELS setup), which is adiabatically focused and, consequently, absorbed as it propagates into the depth of the groove. Quite surprisingly, the experimental measurements showed even better light absorption than the simulations. Those simulations were based on a completely symmetric groove geometry and (for the most part) normally incident light. However, the grooves are not perfectly symmetric, as discussed in the context of Fig.~\ref{fig:fig1}, nor is the light in the experimental setup a perfect plane wave impinging normal to the surface. Thus, incoming light will, due to inclined propagation and the slight asymmetry of the groove, in practice also couple to the significantly more lossy aGSP mode, the one that we studied here with EELS. In fact, it was already shown in the supplementary information of Ref.~\onlinecite{Sondergaard:2012} that a small inclined angle~($\sim20^\circ$) of the incident light moderately improves the overall absorption, which we ascribe to the excitation of the aGSP mode. It should be stressed that the introduction of small asymmetry entails a weak excitation of aGSP, leaving still the dominant contribution of absorption to the sGSP mode. As a second verification of the important role of aGSPs in absorption of light in asymmetric configurations, we have performed light scattering simulations of slightly asymmetric grooves with normal incident light (see Supplementary Figure~3), which confirm the increased absorption relative to the symmetric configuration. In fact, significant absorption increases of $\sim 40$\% in a wide range of energies are observed. This important discovery opens up opportunities to improve the plasmonic black metal effect by deliberately introducing the optimal geometric asymmetry to the structure.

We have reported the first application of EELS to the characterization of extremely confined GSP modes excited by electrons in nanometer-wide gaps. Using ultra-sharp convex grooves in gold, we have recorded the EELS data with high spatial ($<1$~nm) and energy ($\sim 0.15$~eV) resolution throughout the whole groove cross section. Both experimental and theoretical EELS data have revealed resonance behavior associated with the excitation of the aGSP mode for extremely small gap widths, down to $\sim 5$~nm. We have further related the excitation of this mode, featuring very strong absorption, to the experimental results obtained in the study devoted to the phenomenon of plasmonic black gold, in which very efficient and broadband light absorption by ultra-sharp convex grooves has been observed~\cite{Sondergaard:2012}. It has been argued that a part of the light absorption in the grooves can be ascribed to the aGSP excitation due to fabrication induced asymmetry. Finally, it should be stressed that the aGSP excitation should be expected to occur at any asymmetric junction/bend of GSP-based (gap or slot) plasmonic waveguides typically employed in various plasmonic circuits~\cite{Cai:2010}. The aGSP mode absorption thereby represents an additional (efficient) channel of energy dissipation that should be taken into account in the design of plasmonic nanophotonic circuits and devices.

\emph{Acknowledgments.} The Center for Nanostructured Graphene is sponsored by the Danish National Research Foundation, Project DNRF58. The A. P. M{\o}ller and Chastine Mc-Kinney M{\o}ller Foundation is gratefully acknowledged for the contribution toward the establishment of the Center for Electron Nanoscopy. N.S. acknowledges financial support by a Lundbeck Foundation Grant Nbr. R95-A10663.

\end{document}